\documentclass[conference]{IEEEtran}
\IEEEoverridecommandlockouts
\usepackage{cite}
\usepackage{amsmath,amssymb,amsfonts}
\usepackage{textcomp}
\IEEEoverridecommandlockouts
\usepackage[section]{placeins}
\usepackage{algorithmic}
\usepackage{graphicx}
\usepackage{amsmath}
\usepackage{float}

\usepackage{multirow}
\usepackage[utf8]{inputenc}
\usepackage{hhline}
\usepackage{booktabs}
\usepackage{tikz}
\usepackage{subcaption}
\usepackage{colortbl}
\usepackage{xcolor}
\usepackage{hyperref}
\usepackage{amsmath}

\def\BibTeX{{\rm B\kern-.05em{\sc i\kern-.025em b}\kern-.08em
    T\kern-.1667em\lower.7ex\hbox{E}\kern-.125emX}}
\usepackage{cite}
\def\BibTeX{{\rm B\kern-.05em{\sc i\kern-.025em b}\kern-.08em
    T\kern-.1667em\lower.7ex\hbox{E}\kern-.125emX}}
\begin{document}

\title{Two-Stage Triplet Loss Training with Curriculum
Augmentation for Audio-Visual Retrieval}

\author{
\IEEEauthorblockN{Donghuo Zeng}
\IEEEauthorblockA{\textit{KDDI Research, Inc.} \\
Saitama, Japan \\
do-zeng@kddi-research.jp
}
\and
\IEEEauthorblockN{Kazushi Ikeda}
\IEEEauthorblockA{\textit{KDDI Research, Inc.} \\
Saitama, Japan \\
kz-ikeda@kddi-research.jp}
}

\maketitle

\begin{abstract}
The cross-modal retrieval model leverages the potential of triple loss optimization to learn robust embedding spaces. However, existing methods often train these models in a singular pass, overlooking the distinction between semi-hard and hard triples in the optimization process. The oversight of not distinguishing between semi-hard and hard triples leads to suboptimal model performance. In this paper, we introduce a novel approach rooted in curriculum learning to address this problem. We propose a two-stage training paradigm that guides the model's learning process from semi-hard to hard triplets. In the first stage, the model is trained with a set of semi-hard triplets,  starting from a low-loss base. Subsequently, in the second stage, we augment the embeddings using an interpolation technique. This process identifies potential hard negatives, alleviating issues arising from high-loss functions due to a scarcity of hard triples. Our approach then applies hard triplet mining in the augmented embedding space to further optimize the model. Extensive experimental results conducted on two audio-visual datasets show a significant improvement of approximately 9.8\% in terms of average Mean Average Precision (MAP) over the current state-of-the-art method, MSNSCA, for the Audio-Visual Cross-Modal Retrieval (AV-CMR) task on the AVE dataset, indicating the effectiveness of our proposed method.
\end{abstract}


\section{Introduction}
Cross-modal retrieval~\cite{wang2017adversarial, zhen2019deep, pmlr_v139_radford21a, zeng2023learning}, which involves searching for multimedia data (such as images, audio, or video) based on queries from a different modality, has garnered significant attention due to its wide-ranging applications in fields like multimedia retrieval~\cite{zeng2020musictm, zeng2020deep}, generation systems~\cite{zhou2018visual}. A crucial component of cross-modal retrieval models lies in the learning of robust embedding spaces~\cite{wang2017adversarial, zhen2019deep, pmlr_v139_radford21a, zeng2023learning}, where the embeddings of different modalities are mapped to a common space for effective similarity measurements.

While the potential of triple loss optimization~\cite{schroff2015facenet} in enhancing the quality of these embeddings has been acknowledged, a critical oversight in existing methods lies in their singular pass training approach. This approach tends to disregard the distinction between semi-hard and hard triplets during the optimization process (a triplet consists of an anchor, a positive, and a negative, and the distinction depends on the negative samples). Consequently, it leads to suboptimal model performance and inhibits the model's ability to effectively handle challenging retrieval scenarios such as audio-visual tasks~\cite{zeng2018audio, zeng2020deep}.

\begin{figure}[t]
\centering
\includegraphics[width=0.48\textwidth]{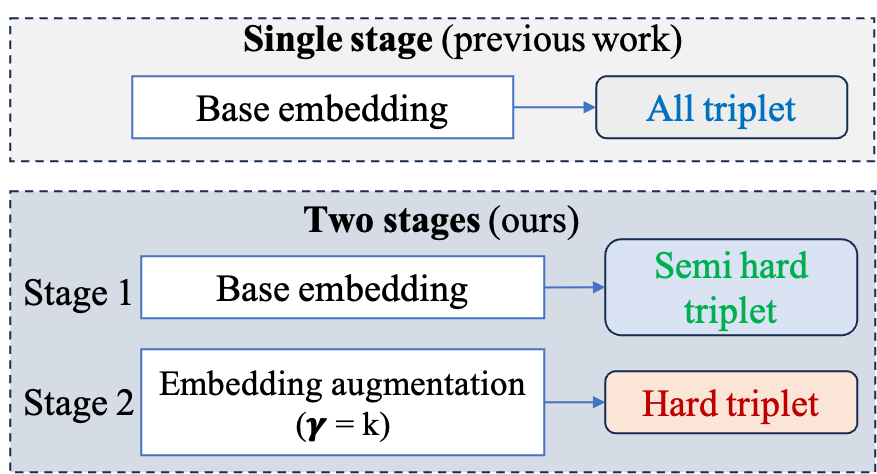}
\caption{This schematic illustrates the comparison between existing single-stage training and our proposed two-stage training approach. In Stage 1, our model is trained using semi-hard triplets. In Stage 2, embedding augmentation techniques are employed to selectively identify hard triplets for further model refinement. The parameter $\gamma$ governs the number of augmented points introduced.}
\label{fig:overall}
\end{figure}
In this paper, we present a novel approach rooted in curriculum learning~\cite{bengio2009curriculum} to bridge this gap and improve the performance of cross-modal retrieval models, shown in Fig.~\ref{fig:overall}. Our method revolves around a two-stage training paradigm designed to guide the model's learning process from semi-hard to hard triplets. In the initial phase, the model is trained with a curated set of semi-hard triplets, starting from a low-loss base. This phase provides a strong foundation for subsequent optimization.

Subsequently, in the second phase, we introduce an augmentation technique used for cross-modal learning that enriches the embeddings through an interpolation process. This augmentation process identifies potential hard triplets, effectively mitigating issues stemming from high-loss functions, particularly in cases where hard triples are scarce. We then engage hard triplet mining within the augmented embedding space to further refine the model's performance.

In summary, our contributions are as follows:
\begin{itemize}
    \item Our proposed two-stage triplet loss training strategy effectively addresses the contribution disparity of hard negative samples during model training, by transitioning from semi-hard triplets to hard triplets, ensuring a more stable and robust training process.
    \item By employing data augmentation through the interpolation method in the second stage, we identify potential hard triplets, thus mitigating the risk of training a biased model.
    \item Our proposed method is evaluated through extensive experiments conducted on two prominent audio-visual datasets. The results demonstrate a substantial improvement of approximately 9.8\% in terms of average Mean Average Precision (MAP) over the current state-of-the-art method, MSNSCA, for the Audio-Visual Cross-Modal Retrieval (AV-CMR) task on the AVE dataset, and attained the second-best performance on the VEGAS dataset. These outcomes affirm that our two-stage triplet loss training coupled with curriculum augmentation effectively optimizes the model training process for cross-modal representation learning.
\end{itemize}

In the following sections, we will provide detailed related works to explain the difference in Section~\ref{related}. In Section~\ref{apporach}, we describe our proposed method, including embedding augmentation and a two-stage training paradigm. Section~\ref{experiment} presents the experimental setup, results, and comparative analysis, further affirming the effectiveness of our proposed approach. Finally, we conclude with a discussion of the implications and potential future directions for this research in Section~\ref{conclusion}.

\section{Related Work}\label{related}
In this section, we introduce both cross-modal retrieval methods and the emerging paradigm of curriculum learning.

\subsection{Cross-modal Retrieval}
The core of cross-modal retrieval is the subspace learning technique, which aims at bridging the disparate domains of heterogeneous modalities. Canonical Correlation Analysis (CCA)~\cite{akaho2006kernel} stands as a longstanding classical method for two sets of variables. It orchestrates a statistical alignment between variable sets through basis vectors. It achieves enduring significance by optimizing the correlation between linear projections in a shared space. Approaches like Kernel CCA (KCCA)~\cite{akaho2006kernel} venture into complex mappings. KCCA uses the "kernel trick" to project cross-modal data into high-dimensional spaces. However, it contends with the perennial challenge of kernel selection. Deep Canonical Correlation Analysis (DCCA)~\cite{andrew2013deep} pioneers the integration of deep learning techniques to uncover complex, nonlinear relationships in data. Building upon the foundation of CCA, DCCA leverages neural network architectures to capture intricate data connections. Taking this a step further, Cluster-based CCA (C-CCA)~\cite{rasiwasia2014cluster} integrates label information with intermodal data clustering. This process encourages stronger connections within data clusters and extracts distinct labels within a shared subspace. The combination of C-CCA and DCCA gives rise to Category-based Deep CCA (C-DCCA)~\cite{yu2018category, zeng2018audio}. It guides cross-modal data through deep neural networks to produce highly correlated points within each cluster. Among these accomplishments, the triplet neural network with cluster CCA (TNN-C-CCA)~\cite{zeng2020deep} achieves a culmination of C-CCA and C-DCCA fortified by audio-visual ranking loss. It transforms label information into a matrix, capturing both similar and dissimilar correlations, emerging as a benchmark on the VEGAS dataset.

Recent advancements take advantage of some popular deep learning models with rank loss. The Adversarial Cross-Modal Retrieval (ACMR) method~\cite{wang2017adversarial} leverages adversarial learning to construct a robust common subspace, enabling the generation of modality-invariant features through triplet loss. This process effectively reduces the representation gap among items from various modalities sharing identical semantic labels, thereby amplifying the model's discriminative capabilities. 
Deep supervised cross-modal retrieval (DSCMR~\cite{zhen2019deep} strengthens supervised learning by integrating a label subspace with the common feature subspace. This two-subspace learning approach effectively preserves discrimination among samples from distinct semantic categories and mitigates cross-modal discrepancies.

However, these models that encompass CCA-based approaches and deep learning-based methods have been traditionally trained in a singular manner. The potential benefits of training these models with curriculum learning have not been sufficiently explored, leaving room for optimization.

\subsection{Curriculum Learning in Cross-Modal Retrieval}
Conventional neural network training sampled random mini-batches from the entire training dataset. This study~\cite{hacohen2019power} investigates the effect of curriculum learning, which employs non-uniform mini-batch sampling. It achieves this by organizing training examples based on difficulty and generating a progression of mini-batches with increasing levels of difficulty. The work~\cite{Huang_2020_CVPR} offers a nuanced training strategy by incorporating curriculum learning into the loss function and dynamically adjusts the importance of easy and hard samples at different stages of training. Initially, it places emphasis on easy samples and gradually shifts towards prioritizing harder ones as training advances. The work~\cite{bian2021contrastive} proposes a curriculum learning strategy that guides contrastive learning through a progressive, easy-to-difficult learning process. They leverage the augmented data sequence and set them in an easy-to-difficult order. Then, conduct contrastive learning via the elaborately designed curriculum. Curriculum learning is also employed in a two-step process in the study of the works~\cite{liu2022competence}. Firstly, it evaluates the difficulty level of each training instance and assesses the current model's proficiency. Then, based on this assessment, it strategically selects an appropriate batch of training instances, ensuring a progressive, step-by-step exposure to the training data.

Although the utilization of this specific loss function for curriculum learning has limited attention and reports, the selection of triplets significantly influences model performance in cross-modal retrieval tasks. This paper studies the application of curriculum learning for the incremental training of various types of triplets.

\begin{figure*}[htbp]
\centering
\includegraphics[width=0.95\textwidth]{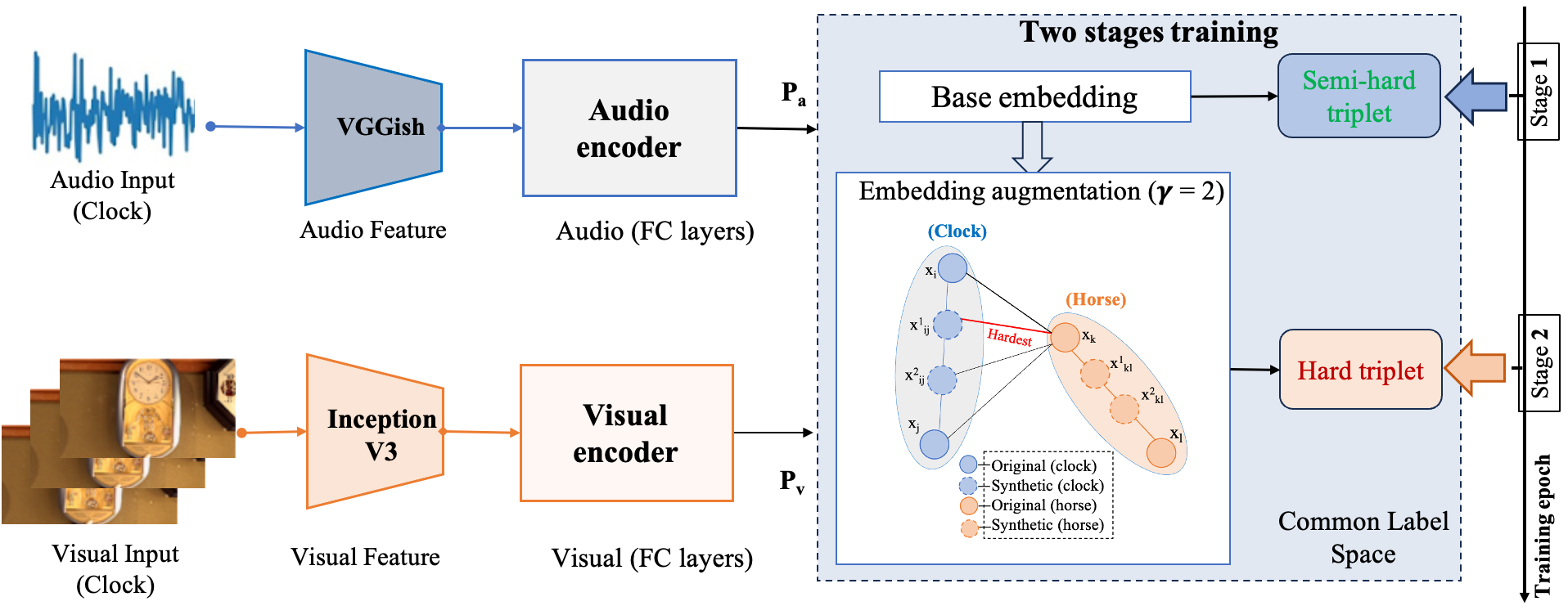}
\caption{This diagram illustrates the key components of our architecture. Initially, audio and visual representations ($a_{i}$ and $v_{i}$) extracted by pre-trained Vggish and InceptionV3 models are projected into a common label space. Subsequently, a two-stage triplet loss with curriculum augmentation is applied for learning correlations between predicted audio and visual label features. During the initial stages of model training (Stage 1), only semi-hard triplets are selected for training. Following a set number of training epochs, we transition to Stage 2, where $\gamma$ samples are interpolated within $x_{i}$ and $x_{j}$ (from the same modality) to augment the training set, $x_{i} \in \{a_{i}, v_{i}\}$, $x_{j} \in \{a_{j}, v_{j}\}$.}
\label{fig:arch}
\end{figure*}

\section{The Proposed Method}~\label{apporach}
\subsection{Problem setting}
Consider a set of n pairs, each consisting of audio and visual elements, denoted as $\Delta = \{(a_{i}, v_{i})\}_{i=1}^{n}$. Here, $a_{i}$ stands for a 128-dimensional audio input, and $v_{i}$ represents a 1024-dimensional visual input for the $i$-th instance. Importantly, each $(a_{i}, v_{i})$ pair carries a label $Y_{i}$ from a set of labels, where the total labels are denoted by  $C$. In the context of cross-modal retrieval, a sample from one modality along with its label is given as a query. The system then creates a ranked list of all samples in the database from the other modality. This ranking is determined by assessing the similarity between the query and each database sample. Notably, samples sharing the same label as the query are given a higher similarity score than those with different labels. This pivotal distinction ensures the retrieval of contextually relevant samples from the other modality within the database in response to the query.

To facilitate the direct measurement of audio and visual inputs, we undertake a projection into a shared subspace. This is achieved by mapping the projected features into label space, denoted as $f(a_{i})$ and $g(v_{i})$ for audio and visual transforms, respectively. The primary objective is to enhance the discriminative $f(a_{i})$ and $g(v_{i})$, while also ensuring that the resulting output distributions retain a high level of semantic distinctiveness and preserve the inherent correlation between the two modalities.

\subsection{Embedding Augmentation} \label{embedding_augmentation}
Embedding augmentation is essential as it enriches the diversity and quality of data points, addressing the challenge posed by limited data, which, in turn, enhances the model's ability to discover potential triplets and improve overall performance in cross-modal retrieval tasks.

Inspired by the approach proposed in the work~\cite{ko2020embedding}, we employ synthetic point generation within the same class in the embedding space to enrich information for the triplet loss. Furthermore, we implement hard triplets mining to focus on the most informative feature representations. Specifically, the embedding augmentation utilizes linear interpolation between two feature points, producing synthetic points that internally divide into $\gamma$+1 equal parts. For a given pair of feature points {$x_{i}$, $x_{j}$} belonging to the same event category in the embedding space, the computation of each synthetic point can be calculated as follows:

\begin{equation} 
\begin{split}
    x^{k}_{ij} &=\frac{k * x_{i} + (\gamma-k)*x_{j}}{\gamma}; \\ 
    S_{ij} &= \{x^{1}_{ij}, x^{2}_{ij}, ..., x^{\gamma}_{ij}\}
\end{split}
\end{equation}

where $\gamma$ is the number of synthetic data points, $S_{ij}$ is a set of generated data points. For each synthetic point, we use $L2$-normalization consistent with the approach detailed in the work~\cite{ko2020embedding} for all subsequent triplet loss computations.

\begin{figure}[ht]
\centering
\includegraphics[width=0.45\textwidth]{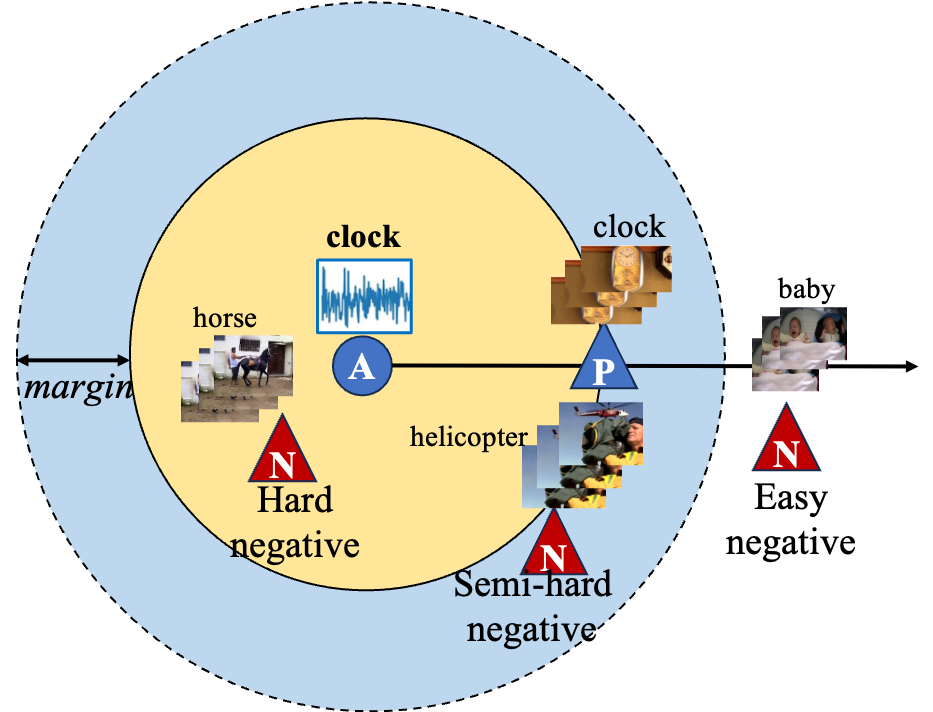}
\caption{The categorization of triplets is determined based on the criteria defined by the triplet loss function. Anchor (A) from the audio, for example, a positive (P) sample from the visual, both labeled with the same class 'clock'. Negative (N) sample from the visual, labeled with a different class. The triplet loss ensures that the distance from A to N is greater by a pre-defined margin (hyperparameter) than the distance from A to P. Negative samples can fall into three possible regions: white, light blue, and yellow, categorizing them as easy, semi-hard, and hard negatives.}
\label{fig:triplet}
\end{figure}
\subsection{Two-Stage Triplet Loss Training}
The effectiveness of training strategies is important in achieving robust performance in cross-modal retrieval tasks. In this subsection, we introduce a two-stage triplet loss training strategy, incorporating the concept of curriculum learning.

The training of the triplet loss function typically categorizes triples into three distinct categories, , as seen in Fig.~\ref{fig:triplet}, and subsequently selects the most informative ones for training. The triplet categories are determined by the following formula:
\begin{equation} 
\begin{split}
L_{triplet} = \sum_{i=1}^{N} \max \left(0, d(a_i, p_i) - d(a_i, n_i) + margin \right)
\end{split}
\end{equation}
where $margin$ is a hyperparameter to govern the degree of separation between positive and negative samples, $d(\cdot)$ is the distance function. The three triplet categories are defined as follows: (a) easy triplet; (b) semi-hard triplet; and (c) hard triplet.
\begin{equation}
\begin{split}
\text{(a)} \quad & d(a_i, n_i) > d(a_i, p_i) + margin \\
\text{(b)} \quad & d(a_i, n_i) < d(a_i, p_i) \\
\text{(c)} \quad & d(a_i, p_i) < d(a_i, n_i) < d(a_i, p_i) + margin
\end{split}
\end{equation}

We employ two kinds of informative triplets to achieve two-stage training processes to enhance the discriminative capabilities of our method, seen in Fig.~\ref{fig:arch}. In the first stage, the model is exposed to all semi-hard triplets, providing a foundational understanding of the data relationships in the common label space. The second stage introduces a crucial refinement step by applying embedding augmentation methods (subsection~\ref{embedding_augmentation}) to selectively identify and incorporate potential triplets for further training. This two-stage approach significantly boosts the model's ability to capture nuanced cross-modal associations. Our final loss function of the proposed method for each stage is as follows.
\begin{equation}
   Loss = Loss_{label} + Loss_{triplet}
    \label{equ:overall_formula}
\end{equation}
where $Loss_{label}$ is the same as the work~\cite{zeng2022complete} to map the feature into common label space. The selected triplets of $Loss_{triplet}$ are semi-hard triplets for stage 1 and hard triplets for stage 2. Ultimately, the optimization of the final loss function is executed using the stochastic gradient descent (SGD) algorithm.

\section{Experiment}\label{experiment}
In this section, we assess the performance of our proposed AV-CMR model by conducting a comprehensive comparison with established state-of-the-art methods. Additionally, we conduct ablation studies, dissecting each facet of our method to gain deeper insights into its effectiveness under varying conditions.
\subsection{Dataset and Metrics}

We achieved success in the AV-CMR task by assuming that the audio and visual modalities share identical semantic information. As a result, we select video datasets containing audio-visual trajectories and ensure that both trajectories are labeled identically on the time series. We select two special video datasets (VEGAS~\cite{zhou2018visual, zeng2023learning} and AVE~\cite{tian2018audio, zeng2023learning}) wherein the audio-visual sequences underwent a thorough process of label double-checking, thereby ensuring uniform labeling across all frames within both modalities. Furthermore, we employ identical data partitioning for training and testing sets, as well as the same approach for feature extraction as the referenced work~\cite{zeng2023learning, zeng2022complete}. 

Furthermore, we employ identical data partitioning for training and testing sets, as well as the same approach for feature extraction as the referenced work~\cite{zeng2023learning, zeng2022complete}. 

For model evaluation metrics, we adopt mean average precision (MAP) for both audio-to-visual and visual-to-audio two-directional retrievals. The MAP functions as an evaluative metric to gauge the effectiveness of models in AV-CMR. 

\begin{table*}[!t]
\caption{Comparison of MAP scores between our proposed approach and state-of-the-art methods. The highest MAP scores are denoted in bold, and the second-highest scores are underlined for emphasis.}
\begin{center}
\scalebox{1.2}{%
\begin{tabular}{|c|c|c|c|c|c|c|}
    \hline
   \multirow{2}{*}{\textbf{Methods}} & \multicolumn{3}{c|}{\textbf{AVE Dataset}} & \multicolumn{3}{c|}{\textbf{VEGAS Dataset}} \\
    \cline{2-7}
  & audio$\rightarrow$visual
  & visual$\rightarrow$audio
  & Average
  & audio$\rightarrow$visual
  & visual$\rightarrow$audio
  & Average
   \tabularnewline 
\hline 
   Random
    & 0.127 & 0.124 & 0.126 & 0.110 & 0.109 & 0.109 \\ \hline
   CCA~\cite{hardoon2004canonical} 
     & 0.190 & 0.189 & 0.190 & 0.332 & 0.327  & 0.330 \\ \hline
   KCCA~\cite{akaho2006kernel}
     & 0.133 & 0.135 & 0.134 & 0.288 & 0.273  & 0.281 \\ \hline
   DCCA~\cite{andrew2013deep}
     & 0.221 & 0.223 & 0.222 & 0.478 & 0.457  & 0.468\\ \hline
   C-CCA~\cite{rasiwasia2014cluster}
    & 0.228 & 0.226 & 0.227 & 0.711 & 0.704  & 0.708 \\ \hline
   C-DCCA~\cite{yu2018category, zeng2018audio}
     & 0.230 & 0.227 & 0.229 & 0.722 & 0.716  & 0.719\\ 
    \hline
    TNN-C-CCA~\cite{zeng2020deep}
     & 0.253 & 0.258 & 0.256 & 0.751 & 0.738 & 0.745\\ \hline
   ACMR~\cite{wang2017adversarial} 
    &  0.162   & 0.159   & 0.161 & 0.465 & 0.442 & 0.454 \\ \hline
   DSCMR~\cite{zhen2019deep}
     & 0.314 & 0.256 & 0.285 & 0.732 & 0.721 & 0.727\\ \hline
   CLIP~\cite{pmlr_v139_radford21a}
     & 0.129 & 0.161 & 0.145 & 0.473 & 0.617 & 0.545\\ \hline
   BiC-Net~\cite{hou2021bicnet}
    & 0.188 & 0.187 & 0.188 & 0.680 & 0.653 & 0.667\\ \hline
   DCIL~\cite{zheng2020dual}
    & 0.244 & 0.213 & 0.228 & 0.726 & 0.722 & 0.724 \\ \hline
   CCTL~\cite{zeng2022complete}
    & 0.328 & 0.267 & 0.298 & 0.766 & 0.765 & 0.766\\ \hline
   EICS~\cite{zeng2023learning}   
   &\underline{0.337} & 0.279 & 0.308 & 0.797 & 0.779 & 0.788 \\ \hline
   VideoAdviser~\cite{wang2023videoadviser} 
   &- &- &-& \underline{0.825} & 0.819 & 0.822 \\ \hline
   MSNSCA~\cite{zhang2023multi}  
   & 0.323 &\underline{0.343} &\underline{0.333} &\textbf{0.866} &\textbf{0.865} &\textbf{0.866}\\ \hline
    \textit{\textbf{Ours}} 
    & \textbf{0.410} &  \textbf{0.451}  &  \textbf{0.431} &  0.822  &  \underline{0.838}  &  \underline{0.830}  \\
\hline
\end{tabular}}
\label{tab:result}
\end{center}
\end{table*}

\subsection{Implementation Details}
Our model architecture consists of three fully connected (FC) layers followed by a label prediction layer for both audio and visual inputs. Each FC layer is configured with 1024 hidden units. Predicted categories and pre-defined categories are set at 10 for VEGAS and 15 for AVE. Training was conducted over 1000 epochs with batch sizes of 400 for both VEGAS and AVE.

We employed PyTorch as our framework on Ubuntu Linux 22.04.2, utilizing an NVIDIA GeForce 3080 (10G). For training, we used the Adam optimizer~\cite{kingma2014adam} with default parameters and a learning rate of 0.0001.

\subsection{Comparison with Existing AV-CMR Approaches}
To assess the effectiveness of our proposed method, a curriculum-based approach to two-stage triplet loss optimization from semi-hard to hard triplet, we perform experiments comparing it against a range of methods including CCA-based methods and state-of-the-art deep learning-based approaches. The CCA-based methods encompass CCA~\cite{hardoon2004canonical}, KCCA~\cite{akaho2006kernel}, DCCA~\cite{andrew2013deep}, C-CCA~\cite{rasiwasia2014cluster}, C-DCCA~\cite{yu2018category}, and TNN-C-CCA~\cite{zeng2020deep}, all designed to discover a shared space enabling the computation of linear or nonlinear projections for two sets of variables while optimizing their mutual correlations.

The state-of-the-art deep learning-based approaches include ACMR~\cite{wang2017adversarial}, DSCMR~\cite{zhen2019deep}, CLIP~\cite{pmlr_v139_radford21a}, BiC-Net~\cite{hou2021bicnet}, DCIL~\cite{zheng2020dual}, CCTL~\cite{zeng2022complete}, EICS~\cite{zeng2023learning}, VideoAdviser~\cite{wang2023videoadviser}, and MSNSCA~\cite{zhang2023multi}. ACMR employs supervised adversarial learning for cross-modal representation learning. DSCMR model is an advanced supervised cross-modal retrieval approach that excels in generating highly discriminative representations.  CCTL model captures all possible cross-triplet losses to learn a common subspace for audio-visual learning. EICS model learns two distinct subspaces to capture modality-invariant and modality-specific features across modalities. Videovisor model apply knowledge distillation techniques to distill pre-trained CLIP model~\cite{pmlr_v139_radford21a} to advise the cross-modal retrieval model. MSNSCA model applies multi-scale networks with shared cross-attention for optimizing the audio-visual relationships. We also include three recently developed models that first achieved on other cross-modal retrievals: CLIP~\cite{pmlr_v139_radford21a}, BiC-Net~\cite{hou2021bicnet}, and DCIL~\cite{zheng2020dual}, in our experiments.

\begin{figure*}[h]
\centering
\includegraphics[width=0.98\textwidth]{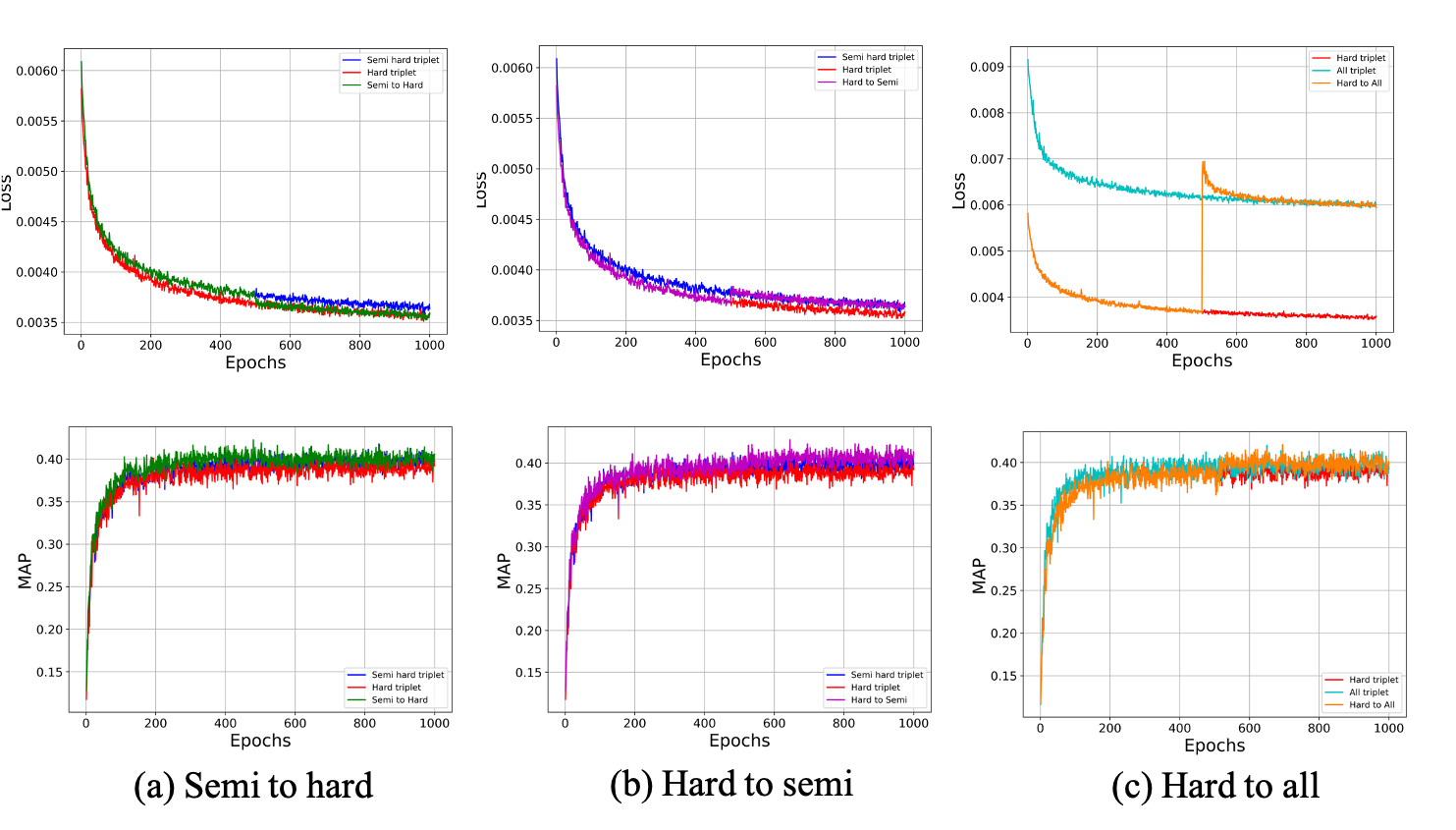}
\caption{The loss values and MAP performance on both the training and test sets on the AVE dataset. It offers a comprehensive comparative examination of different triplet loss variants with their base losses, encompassing group (a) semi to hard triplet loss, group (b) hard to semi triplet loss, and group (c) hard to all triplet loss.}
\label{fig:cl}
\end{figure*}

Table~\ref{tab:result} displays the MAP scores achieved by our method alongside other methods, evaluated on both the VEGAS and AVE datasets. Our method demonstrates strong overall performance, notably surpassing prior state-of-the-art methodologies on the AVE dataset and achieving the second-best performance on the VEGAS dataset. The analysis and summary of the overall performance in comparison with others are presented as follows:
\begin{itemize}
    \item Our approach exhibits a significant improvement, achieving an enhancement of 8.7\% and 10.8\% in MAP for audio-to-visual and visual-to-audio retrieval respectively, coupled with a  9.8\% increase in average MAP on the AVE dataset. Moreover, our method is further confirmed through an evaluation on the VEGAS dataset, demonstrating its effectiveness in both audio-to-visual and visual-to-audio retrieval tasks.

    \item Supervised learning outperforms unsupervised learning due to its utilization of label information, highlighting its critical role of labels in cross-modal representation learning. This insight underscores the effectiveness of mapping features into a common label space, a key factor contributing to our model's superior performance compared to other models operating in a common feature subspace (e.g., C-CCA, C-DCCA, ACMR, DSCMR, EICS, etc.).
    
    \item Incorporating cross-modal representation learning within a common label subspace, and employing rank loss methods like triplet loss (TNN-C-CCA, ACMR) and contrastive loss (CLIP), proves to be a highly effective strategy. Our approach builds upon the CCTL model by implementing a curriculum-guided approach with data augmentation in a two-stage training process, leading to substantial improvements. Additionally, our model boasts a more powerful neural network and larger batch size, further enhancing its performance.
    
    \item When compared with other models utilizing rank loss (TNN-C-CCA, ACMR), particularly triplet loss, our method consistently achieves top-tier results. This success can be attributed not only to the use of a common label space but also to the two-stage training process.
\end{itemize}

\subsection{Ablation Study}
To further investigate our proposed method, we analyze it in terms of three aspects: 1) impact of the curriculum learning; 2) impact of the number of synthesis points; and 3) audio-visual retrieval case study.

\begin{figure}[t]
\centering
\includegraphics[width=0.48\textwidth]{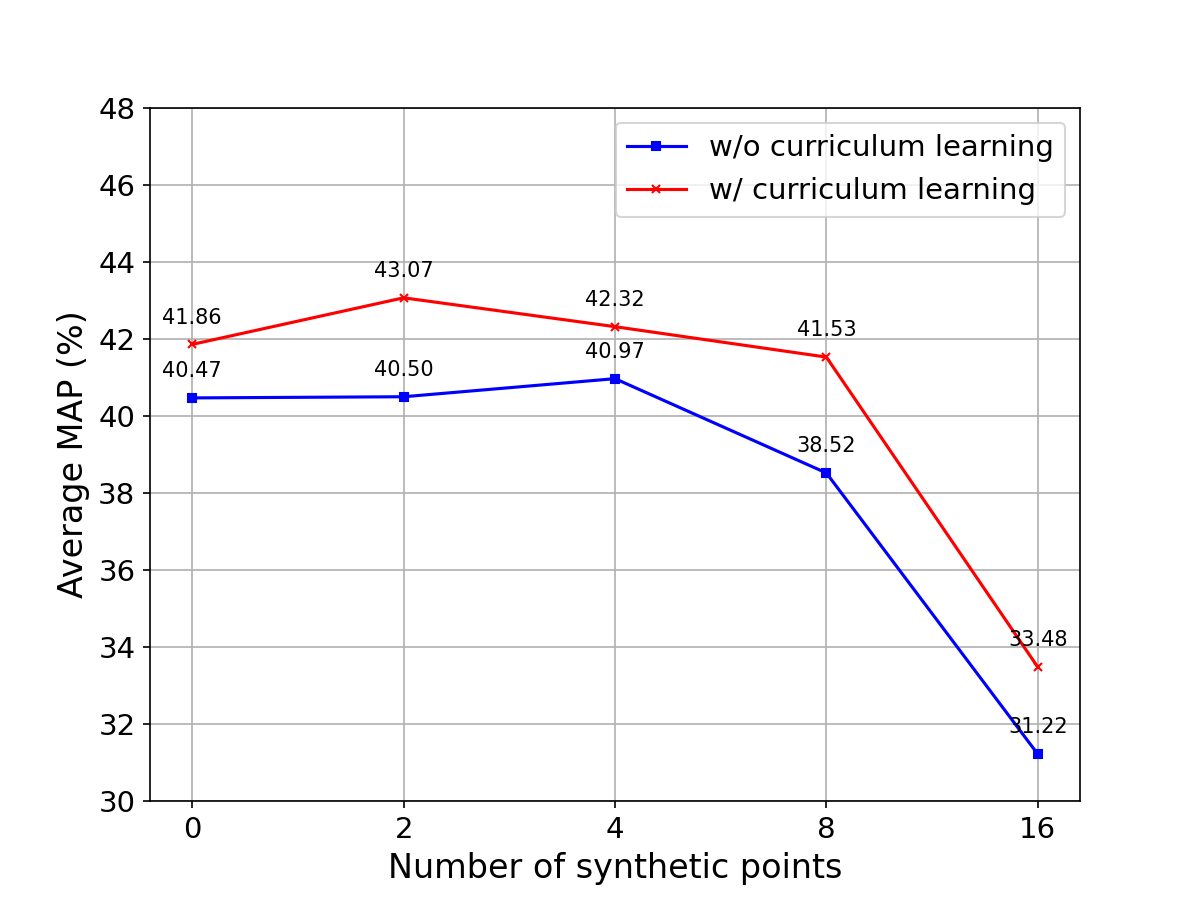}
\caption{Comparison of average MAP (\%) performance on the AVE dataset for models trained with triplet loss and data augmentation. Results are presented with and without the application of curriculum learning.}
\label{fig:syn}
\end{figure}

\subsubsection{Impact of curriculum learning}
In this subsection, we investigate the impact of integrating curriculum learning into our training process. By gradually introducing complexity from semi-hard to hard triplets during training, curriculum learning shows promise in enhancing model performance.

We present the results of our investigation in Fig.~\ref{fig:cl}, where we compare the loss values and MAP performance on both the training and test sets of the AVE dataset. This comprehensive analysis includes three different triplet loss variants: standard triplet loss, triplet loss with augmented synthetic points, and triplet loss combined with curriculum learning.

\begin{figure*}[!t]
\centering
\includegraphics[width=\textwidth]{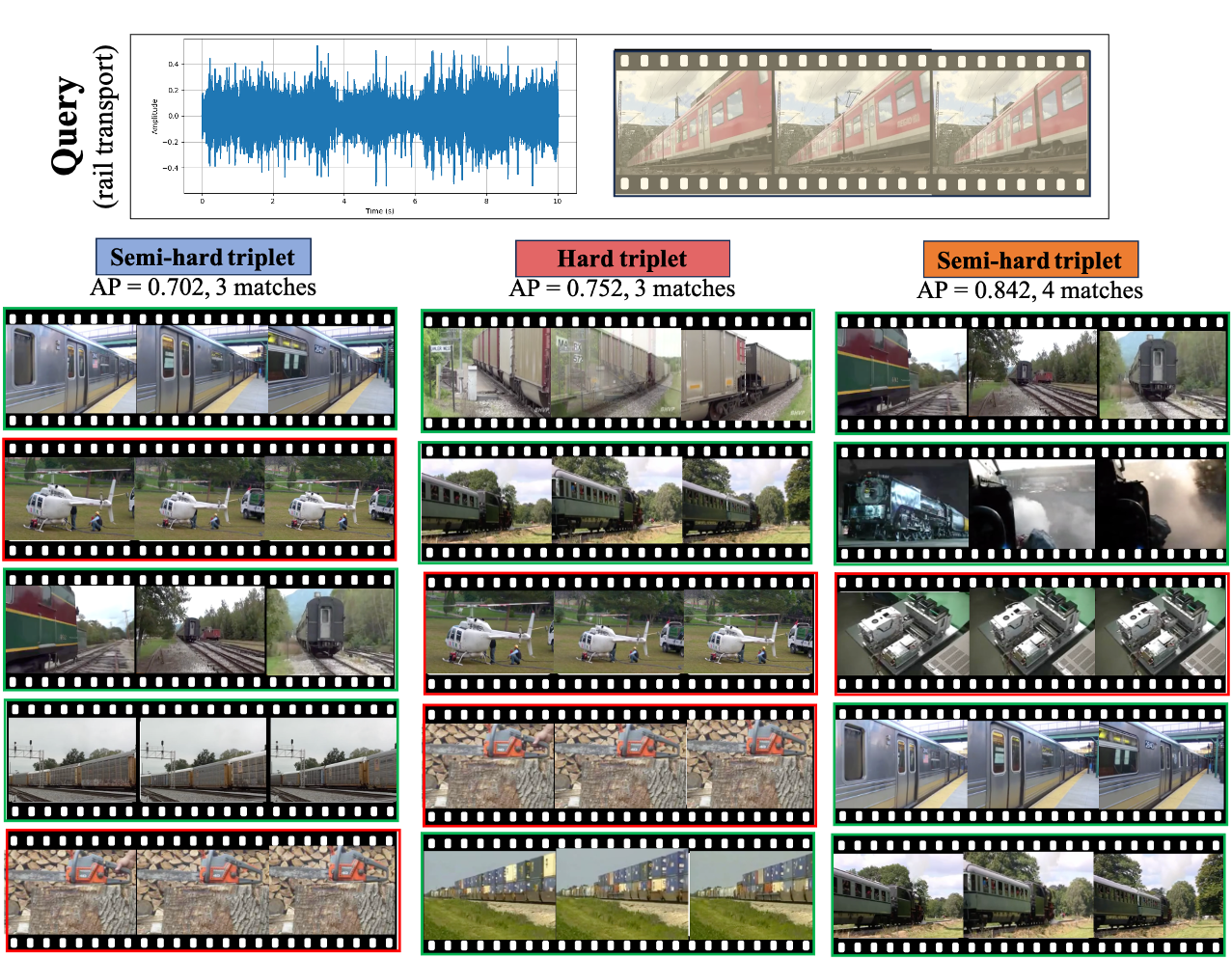}
\caption{Qualitative visual retrieved results: for the given query of a "rail transport" audio from the VEGAS dataset, visual samples for each triplet class are arranged in descending order based on their similarity scores with the query. Visuals enclosed in green boxes represent successful matches, while those in red boxes indicate mismatches.}
\label{fig:retrieval}
\end{figure*}
As depicted in Fig.~\ref{fig:cl}, the impact of curriculum learning is evident in both the loss-epochs and MAP-epochs sub-figures for the training and testing phases. In (a), we observe that the loss curve for the semi-to-hard case initially aligns with the semi-hard loss for the first 500 epochs, and then it begins to follow the trend of hard triplet loss in the last 500 epochs. Additionally, the MAP performance of the semi-to-hard case surpasses that of the other two. Furthermore, (b) and (c) further prove the potential advantages of incorporating curriculum learning into our AV-CMR model training process.

\subsubsection{Impact of the number of synthesis points}
The $\gamma$ in the embedding augmentation is the only hyperparameter in our method. In this subsection, we examine how the number of synthesized points influences the performance of our method. The inclusion of synthetic data points through augmentation techniques is a critical aspect of our approach, and here we explore its effect on the average MAP (\%) performance. 

Fig.~\ref{fig:syn} provides a visual comparison of the average MAP (\%) performance on the AVE dataset for models trained with triplet loss and data augmentation. we compare our method with or without data augmentation in stage 2 during model training.

The results depicted in Fig.~\ref{fig:syn} reveal insightful trends regarding the influence of the number of synthetic points, and it achieves the best result when the number of synthetic points is set as 2 for both. This analysis is crucial in fine-tuning the training process and ensuring optimal performance for our AV-CMR model.

\subsubsection{Audio-visual retrieval case study}
To validate the effectiveness of our proposed method, we conducted a comprehensive case study. In Fig.~\ref{fig:retrieval}, we cherry-pick up the audio with 'rail transport' as a query,  by computing the similarity between the query and each visual to get the rank list. This process was executed using three distinct types of triplets: semi-hard triplets, hard triplets, and semi-to-hard triplets. We use the top five retrieved visual samples from the rank list, each accompanied by its respective label. Based on this truncated ranking list, we computed the Average Precision (AP), denoted as $AP@5$ in our case. We observed that both the semi-hard and hard triplet methods yielded three matching visual samples pertaining to the "rail transport" event. Remarkably, our best model, employing a semi-to-hard triplet strategy, excelled by acquiring four visual samples perfectly aligned with the query event. Through our curriculum-based triplet loss methodology, we achieved a superior $AP@5$ score compared to the two singular triplet loss approaches. Specifically, the curriculum-based triplet loss demonstrated a noteworthy AP@5 score of 0.842, surpassing the semi-hard (AP=0.702) and hard triplet (AP=0.752) methods.

In this case study, we observed that our proposed model exhibits superior retrieval accuracy, corroborating the findings presented in both Table I and Table 1.~\ref{tab:result}.

\section{Conclusion}\label{conclusion}
In this study, we present a novel approach for training cross-modal retrieval models, leveraging the paradigm of curriculum learning. Our two-phase methodology, which systematically advances from semi-hard to hard triplets with embedding augmentation, marks a significant advancement in this field. Through extensive evaluations on benchmark datasets, we have demonstrated the robustness and effectiveness of our approach. Beyond the current work, further enhancements in the curriculum learning strategy, tailored to adapt to the evolving capabilities of the model, hold the promise of yielding even more substantial improvements—an avenue that we envisage for future exploration.

\section*{Acknowledgment}
The first author expresses deep gratitude to Kazushi Ikeda for their invaluable assistance in refining the quality of the paper's writing.

\bibliography{refs}
\bibliographystyle{plain}

\end{document}